\documentstyle[emulateapj,epsfig]{article}
\renewcommand{\baselinestretch}{1.0}
 
\font\smcap=cmcsc10 
\newcommand\hi{{\smcap H$\,$i}}
\newcommand\MHI{$M_{\rm HI}$}
\newcommand\NHI{$N_{\rm HI}$}

\newcommand\kps{km s$^{-1}$}
\newcommand\et{et~al.}			
\newcommand\fsi{${\mit\Sigma}$}
\lefthead{Sansom, Hibbard, \& Schweizer}
\righthead{Gas Content of Fine-Structure E and S0 Galaxies}

\begin{document}
\title{The Cold and Hot Gas Content of Fine-Structure E and S0 Galaxies}

\author{A.~E. Sansom}
\affil{Centre for Astrophysics, University of Central Lancashire, 
Preston PR1 2HE, UK; a.e.sansom@uclan.ac.uk}

\author{J.~E. Hibbard}
\affil{National Radio Astronomy Observatory\altaffilmark{1}, 520 Edgemont 
Road, Charlottesville, VA 22903; jhibbard@nrao.edu}

\medskip
\author{Fran\c cois Schweizer}
\affil{The Observatories of the Carnegie Institution of Washington, 
813 Santa Barbara Street, Pasadena, CA 91101-1292; schweizer@ociw.edu}

\altaffiltext{1}{The National Radio Astronomy Observatory is 
operated by Associated Universities, Inc., under cooperative agreement 
with the National Science Foundation.}

\begin{abstract}

We investigate trends of the cold and hot gas content of early-type
galaxies with the presence of optical morphological peculiarities, as
measured by the fine-structure index \fsi.  \hi\ mapping observations
from the literature are used to track the cold-gas content, and archival
ROSAT PSPC data are used to quantify the hot-gas content.
We find that E and S0 galaxies with a high incidence of optical peculiarities
are exclusively X-ray underluminous and, therefore, deficient in hot gas.
In contrast, more relaxed galaxies with little or no signs of optical
peculiarities span a wide range of X-ray luminosities.  That is, the X-ray 
excess anticorrelates with \fsi. There appears to be
no similar trend of cold-gas content with either fine-structure 
index or X-ray content.  The fact that only apparently relaxed E and S0 
galaxies are strong X-ray emitters is consistent with the hypothesis that 
after strong disturbances such as a merger hot-gas halos build up over a 
time scale of several gigayears. This is consistent with the expected
mass loss from stars.

\end{abstract}


\section{Introduction}
\label{sec:intro}

In the merger hypothesis for elliptical galaxy formation (Toomre \&
Toomre 1972) spiral galaxies coalesce to produce ellipticals.  If this
is a common mechanism for the formation of ellipticals, then it
requires that the cold gas present in the spiral progenitors
(typically $\sim$10$^9$ -- $10^{10} M_{\odot}$) be lost or heated,
since similar quantities of cold gas are not generally detected in
ellipticals (Knapp \et\ 1985a; Lees \et\ 1991; Roberts \et\ 1991). 
On the other hand, elliptical and S0 (i.e., early-type) galaxies are found
to be much more X-ray luminous than spirals at a given blue luminosity 
$L_B$. While there is scatter of two orders of magnitude in the relationship 
between the X-ray luminosity $L_X$ and $L_B$, early-type galaxies follow a 
much steeper relationship than spirals (Fabbiano, Kim \& Trinchieri 1992;
Beuing \et\ 1999). It is therefore of interest to examine the state of the
interstellar medium (ISM) in merger remnants to see whether cold
gas is being converted into other phases. Early-type galaxies showing 
morphological or kinematical peculiarities are suggested to be these
post-merger objects (Schweizer 1986; Schweizer \et\ 1990; 
Bender \& Surma 1992; Schweizer \& Seitzer 1992, hereafter SS92).

Ideally, one would like to trace the evolution of the cold, warm, and
hot gas phases along
a sequence of progressively more evolved merger remnants.  This
requires an observational characteristic that tracks the time elapsed
since merging.  To this end, Schweizer \et\ (1990) introduced a
fine-structure index \fsi\ to quantify the presence of ripples,
jets of luminous matter, boxiness of isophotes, etc.  Ellipticals and
S0 galaxies with larger values of \fsi\ were statistically shown to
have observational characteristics (colours and spectral-line strengths)
consistent with the presence of younger stellar populations,
supporting the idea that they represent more recent mergers in which
starbursts have occurred (Schweizer \et\ 1990; SS92). These galaxies were
suggested to be $\sim$2--10 Gyr old relics of ancient mergers populating
the so-called ``King Gap'' between $\sim 1$ Gyr old remnants and
old ellipticals (I.\ King, quoted in Toomre 1977).

The X-ray properties of a sequence of {\it ongoing} mergers was studied by
Read \& Ponman (1998). They found that, following early increases in the
emission from hot gas, examples of relaxed remnants at $\sim$1.5 Gyrs after
a merger appear relatively devoid of such gas (see also Hibbard \et\ 1994). 
For normal galaxies
of early Hubble type, Bregman \et\ (1992) noted a striking lack of
correlations between the hot and cold interstellar-gas components.
Toward later Hubble types they found a general increase of the cold-to-hot
gas ratio and interpreted this systematic change as evidence that
cold gas is a phenomenon associated with discs, while hot gas is associated
with bulge components.

In this paper we examine the cold and hot atomic gas contents of early-type
galaxies studied by SS92 and which have also been mapped in the 21cm
line of neutral hydrogen (\hi) or observed in X-rays with ROSAT and its
PSPC detector.  By looking for trends with \fsi, we
hope to trace the fate of gas as merger remnants evolve.  Preliminary
work in this area was carried out by Fabbiano \& Schweizer (1995) and
Mackie \& Fabbiano (1997), who showed that three dynamically young
ellipticals with fine structure (NGC 3610, 4125 and 4382) are low X-ray
emitters.  We build upon this earlier work by expanding the number of
galaxies with both fine structure and X-ray content quantified, and by
including information on the cold atomic gas content as well.
Section~\ref{sec:SS92} describes the SS92 sample of galaxies,
while \S~\ref{sec:ROSAT} presents a compilation of archival ROSAT
observations for galaxies from that sample. In
\S~\ref{sec:discussion}, we examine separately and then together the
presence of \hi\ and X-ray emission in the SS92 sample in order
to understand whether galaxies with fine structure are King Gap
objects, and---if so---what we can learn about the fate of cold gas in
disk--disk mergers and the origin of X-ray halos in elliptical
galaxies. Finally, \S~\ref{sec:summary} summarizes our conclusions.

\section{Sample Description}
\label{sec:SS92}

To look at correlations between cold- and hot-gas content and fine structure,
we took the sample of SS92.  This sample consists of 69 E and S0 galaxies
plus the two merger remnants NGC 3921 and NGC 7252.  With the exception
of the latter two objects, all sample galaxies lie north of $\delta$~=
$-$20$^\circ$ and at $|b| > 20^\circ$, have apparent magnitudes
$B_{\rm T}$ $\le$ 13.5, and have recession velocities
$v_0 < 4000$ km s$^{-1}$. The two merger remnants have
$v_0 < 6000$ km s$^{-1}$. Five of the sample galaxies are members of the
Virgo cluster, while all others are either field objects or in loose
groups.  The absolute magnitudes range between $M_B$~= $-$18.4 and $-$21.6,
with a median value of $-$20.3 (for $H_0$~= 75 km s$^{-1}$ Mpc$^{-1}$). 
The merger remnants NGC 3921 and NGC 7252 were included because of their
extremely rich fine structure and evolutionary status as likely
proto-ellipticals (Toomre \& Toomre 1972; Barnes 1994; Hibbard \et\ 1994;
Schweizer 1996).

The fine structure of these galaxies was studied by SS92 on CCD images
obtained in the $R$ passband with the KPNO 0.9 m telescope.
For each galaxy a fine-structure index \fsi\ was derived. This index is
defined as\ \ ${\mit\Sigma} = S + \log(1 + n) + J + B + X$,
where $S$ is a visual estimate of the strength of the most prominent
ripples ($S$~= 0--3), $n$ is the number of detected ripples ($n$~= 0--17),
$J$ is the number of luminous plumes or ``jets'' ($J$~= 0--4), $B$ is a
visual estimate of the maximum boxiness of isophotes ($B$~= 0--3), and
$X$ indicates the absence or presence of an X-structure ($X$~= 0 or 1).
The values of \fsi\ for the present sample of galaxies cover the range\ \
0--10.1. The correlation of \fsi\ with optical colours and line strengths
supports its use as an indicator of ancient mergers (SS92).
High sigma systems are unlikely to have formed from minor
perturbations or internal instabilities, particularly those with tidal
tails or jets indicative of kinematically cold progenitors.
By design, \fsi\ measures four types of fine structure thought to
be caused by mergers and can serve as a rough measure of dynamical youth
or rejuvenation (for detailed review, see Schweizer 1998).

\section{ROSAT Data}
\label{sec:ROSAT}

For the hot-gas content, we first cross-correlated the SS92 list against the
ROSAT PSPC archive (within 20\arcmin\ of the PSPC field centers), which
yielded pointed X-ray observations for 28 of the 71 sample galaxies.
X-ray fluxes were obtained from PSPC images using the broadband 
(0.1--2.4 keV) images from the ROSAT PSPC archive at Leicester university. 
We visually inspected the images to select source and background regions. 
We then used a circular aperture of between 1\arcmin\ and 7\arcmin\ radius
$R$ (depending on source extent) to 
estimate source counts. Background counts in the source detection cell
were estimated using either an annulus or offset areas, depending 
on where other sources lay in the field. To estimate fluxes from these 
source counts we had to adopt an integrated \hi\ column density
(\NHI )  due to our Galaxy. Such column densities were obtained
from the data of Stark \et\ (1992) using a program written by K.\ Arnaud.
We assumed a thermal-source spectrum of solar metallicity and $kT=0.5$ keV 
(typical of many ellipticals, Brown \& Bregman 1998). Using the ROSAT 
Announcement-of-Opportunity document we could
then estimate conversion factors to convert our 
source counts to fluxes (to within 10\%--20\%, allowing for 
a temperatures range of $0.3 < kT < 1.0$ keV). One-sigma errors on fluxes 
are a combination of Poisson errors on source counts, including 
background uncertainty, and estimated errors on the conversion factor. 
These results are presented in Table~\ref{tab:ROSAT}.  The galaxies
NGC 596 and NGC 4283 were undetected, whence $3\sigma$ upper limits 
are given for the source counts and fluxes in these two cases. 

To try to increase the sample size we then looked for galaxies lying
within 20\arcmin\ to 60\arcmin\
of PSPC pointings and found 14 cases, four of which were either
obscured by the PSPC support structure or too close to the edge of the 
field to be measurable. Of the other ten, nine were non-detections and
one (NGC 3065) was detected. The counts for these sources were corrected
for vignetting and are given in Table~\ref{tab:ROSAT} as well.

In addition, we included those sample galaxies that were observed 
as part of the ROSAT All Sky Survey (Beuing \et\ 1999), but
were not among the pointed observations.
This provided a measure of the X-ray content for an additional 12
galaxies from the SS92 sample, although all but one of these (NGC 5982) 
are upper limits.
The galaxies NGC 3377 and NGC 3379, which have ROSAT HRI (but no PSPC)
observations are also listed (from Brown \& Bregman 1998).

Table~\ref{tab:SS92} presents data for all sample galaxies with available
\hi\ mapping and/or ROSAT X-ray observations.  The fine-structure \fsi\ was
taken from SS92.  The total apparent blue magnitudes ($B_{\rm T}$) are from
de Vaucouleurs \et\ (1991, hereafter RC3), and the distances (D) from the 
nearby-galaxies catalogue by Tully (1988) for galaxies within about 40 Mpc 
and from optical velocities given in RC3 for more distant galaxies 
(assuming $H_0=75$ km s$^{-1}$ Mpc$^{-1}$).
Also listed is the logarithm of the ratio of the X-ray 
to blue luminosities, the ``X-ray excess'' $\log(L_X/L_B)$, with $L_X$ in 
ergs $s^{-1}$ and $L_B$ in $L_{B_{\odot}}$.

\begin{figure*}
      \epsfig{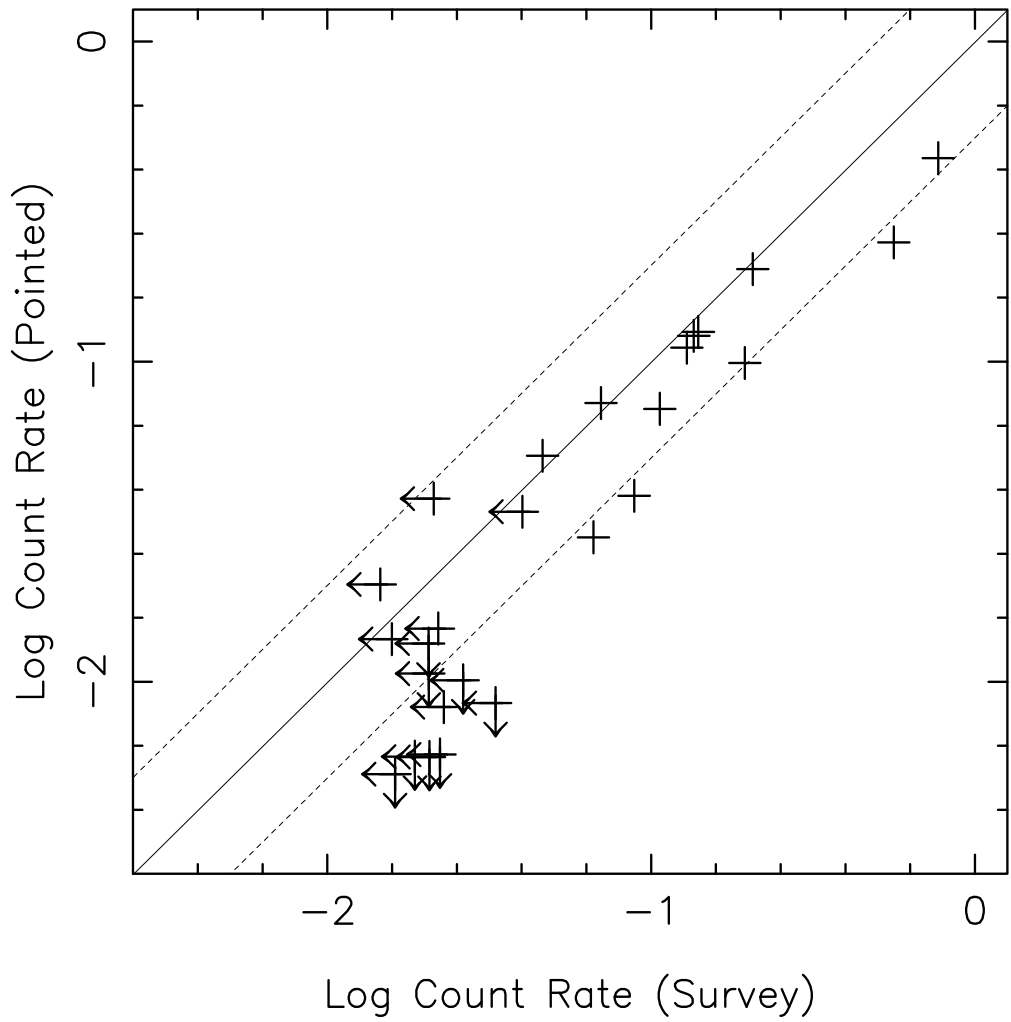}    
\caption{Comparison of count rates (counts s$^{-1}$) measured from the 
ROSAT survey data by Beuing \et\ (1999) with the count rates we have 
measured from archival ROSAT PSPC pointed observation. The solid line 
represents the one-to-one correlation, and the dashed lines are a factor 
of two either side of this. The slightly lower average rates 
measured from the pointed data are less affected by source 
confusion and are, therefore, better estimates of the galaxy emission.
Arrows indicate 3$\sigma$ upper limits to count rates for undetected 
sources.}
\label{fig:Crates}
\end{figure*}

\begin{figure*}
      \epsfig{file=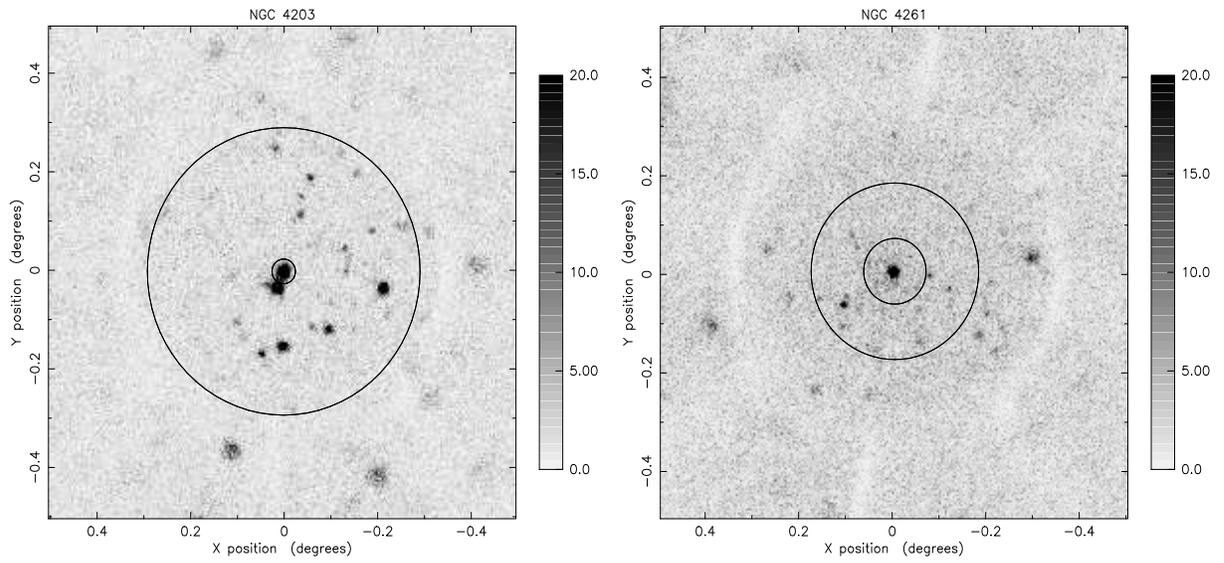,height=18.5cm, width=7.5cm, angle=-90}    
\caption{Examples of two ROSAT pointed images showing a one degree square 
field around the PSPC field center in each case. The galaxies being measured
are NGC 4203 in the left image and NGC 4261 in the right image. 
These images illustrate the 
source confusion affecting the data included in the apertures used on survey 
data by Beuing \et\ 1999 (large circles). The pointed data reduce
this confusion without missing galaxy light (smaller circles). The keys are 
in counts per pixel.}
\label{fig:Images}
\end{figure*}

Several sources appear in both our list of ROSAT PSPC pointed observations 
and in the ROSAT survey data published by Beuing \et\ (1999). As a way of 
checking the reliability of our measurements from the pointed data we 
compared our count rates with those of Beuing et al. The results are
plotted in Fig.~\ref{fig:Crates}. We find that the count rates generally
agree to within a 
factor of two. Where sources are undetected the pointed data provide 
more stringent upper limits. There are a few detections where the rates 
differ by just over a factor of two. These are NGC 7626, which has an 
asymmetric distribution of X-rays and is in the Pegasus I cluster with 
other X-ray emitting galaxies nearby; NGC 3226, which is very close to 
the X-ray bright galaxy NGC 3227; and NGC 4203, which lies only 2\arcmin\
from an apparently unrelated X-ray source of similar brightness. 
In NGC 3226 and NGC 4203, the extraneous sources lie
well within the survey apertures used by Beuing et al.
We have attempted to eliminate their counts from our analysis of the pointed 
data either by using smaller apertures (where this appeared to include all 
the galaxy counts) or by subtracting the extraneous-source counts from the
counts in our galaxy apertures. Survey data were generally obtained with
much shorter exposure times than 
pointed data. The larger point spread functions for the survey data, and for 
the pointed data beyond the central 20\arcmin\ radius of the PSPC, also 
exacerbate the problem of weak source detection.  This is why many more 
galaxies were detected within the central regions of the pointed data 
(see Table~\ref{tab:ROSAT}). The pointed data clearly show that the larger 
radii used by Beuing \et\ include sources separate from the target galaxy.
Figure~\ref{fig:Images} shows two examples.
This effect contributes to the scatter in Fig.~\ref{fig:Crates} and also 
to the systematically lower rates measured from the pointed data. These
rates are less affected by source confusion.

Our computed values of $\log(L_X/L_B)$ agree with published results
within the errors for NGC 3610 and NGC 4125, both also measured from
ROSAT pointed observations by Fabbiano \& Schweizer (1995).

\section{Gas Phase versus Fine Structure}
\label{sec:discussion}

Table~\ref{tab:SS92} shows the compilation of optical, radio, and X-ray
data for the sample galaxies for which these data are available.
In this table, we list all galaxies from SS92 with either \hi\ mapping 
observations or ROSAT observations or both, arranged in order of
decreasing \fsi. This allows us to investigate how the cold- and
hot-gas contents of E and S0 galaxies vary with fine structure.

For the cold-gas content, we used our own data (Hibbard \& Sansom 2000)
and results from a program to map the \hi\ in a sample of peculiar
ellipticals, kindly communicated by J.\ van Gorkom and D.\ Schiminovich (cf.\
Schiminovich \et\ 1994, 1995, 1997; van Gorkom \& Schiminovich
1997).  We also correlated the SS92 sample against the compilation of
\hi\ mapping observations by Martin (1998). Detection limits were 
calculated by taking the quoted 3$\sigma$ noise limits and adopting 
a velocity width of $\Delta v=42$ \kps. While total \hi\ line widths 
may be much wider than this, it is our experience that individual tidal
features have relatively narrow line widths and that this smaller limit is
appropriate for the detection of tidal \hi. We define an ``\hi\ excess''
as $\log(M_{\rm HI}/L_B)$, with the \hi\ mass \MHI\ in solar masses and
$L_B$ in solar luminosities. This allows us to compare the cold- and
hot-gas contents in a similar way.

\begin{figure*}
      \epsfig{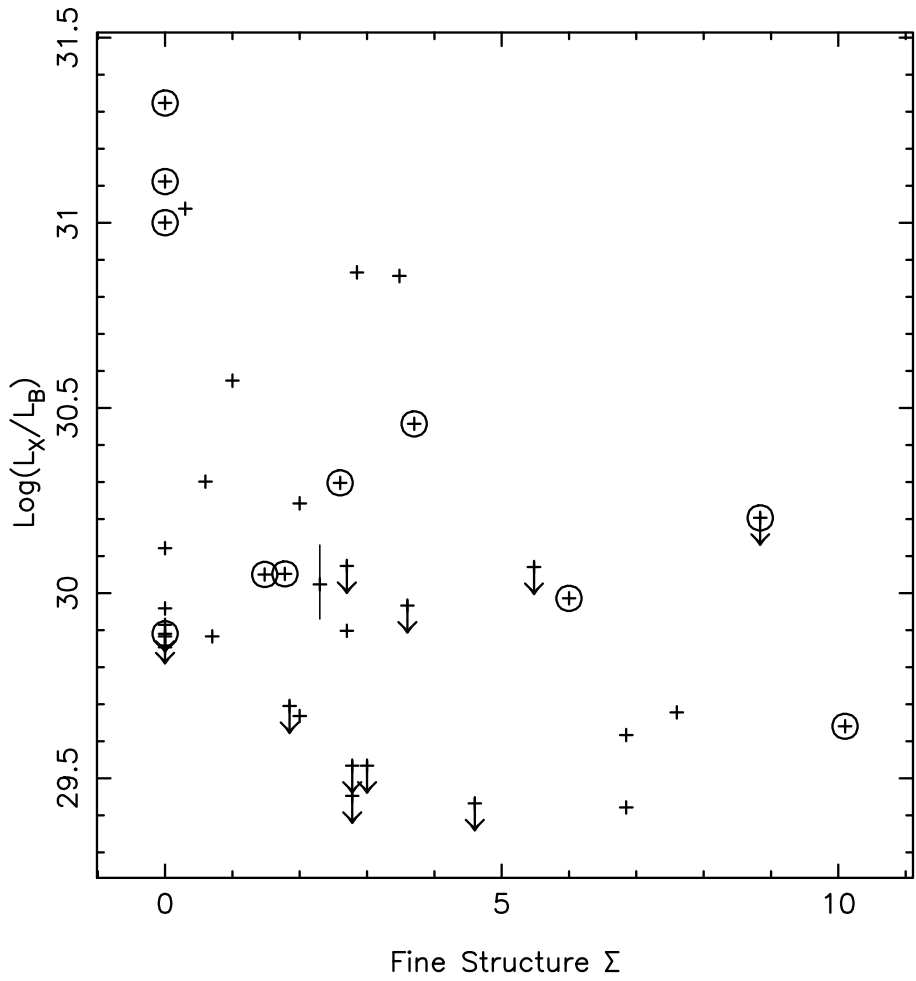}    
\caption{X-ray excess $\log(L_X/L_B)$ (erg s$^{-1}$ $L_{B_{\odot}}^{-1}$)
plotted versus optical fine-structure 
index \fsi\ for 38 galaxies
from the SS92 sample.  The plotted data are from ROSAT PSPC pointed
observations.  Plus signs indicate X-ray detections, while arrows indicate
$3\sigma$ upper limits. Encircled plus points have associated \hi\ 
detections. The vertical bar indicates typical errors.}
\label{fig:LxLb}
\end{figure*}

Figure~\ref{fig:LxLb} shows the X-ray excess plotted against the
fine-structure index for the ROSAT PSPC pointed data.  Circles indicate
\hi\ detections.  From this figure we find a trend for the X-ray
excess to anticorrelate with fine structure. Including only X-ray
detections, the Pearson correlation coefficient is $r=-0.47$, indicating
a clear anticorrelation.  The X-ray excess is large only in early-type
galaxies with low levels of fine structure. Conversely, all galaxies with
high fine-structure index are X-ray weak.
 
\begin{figure*}
      \epsfig{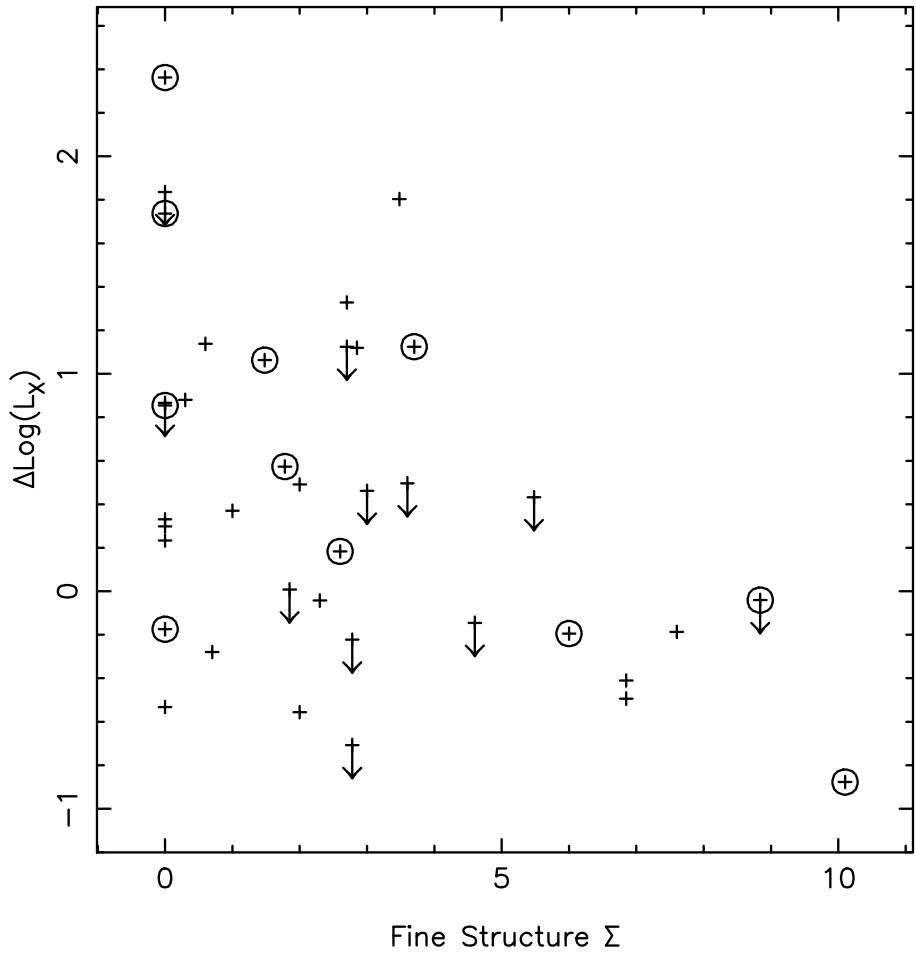}    
\caption{Residuals $\Delta \log(L_X)$ (erg s$^{-1}$) against the mean 
$L_X$--$L_B$ relation
plotted versus fine-structure index \fsi.  The symbols are as in
Fig.~\ref{fig:LxLb}.  The X-ray data plotted are only those obtained from
ROSAT PSPC pointed observations.}
\label{fig:Dx}
\end{figure*}

To investigate this anticorrelation further we looked at how the vertical
scatter in the $L_X$--$L_B$ relation varies with fine structure.
From ROSAT survey data the mean $L_X$--$L_B$ relation found
for early-type galaxies by Beuing \et\ (1999, esp.\ Fig.~8) was 
$$ \log(L_X/{\rm erg\: s}^{-1}) = 2.23 \log(L_B/L_{B_{\odot}}) + 17.02 $$ 
(i.e. $L_X \propto L_B^{2.23}$).  We used this relation to compute
residuals $\Delta\log(L_X)$ against the mean value of $\log(L_X)$ at any
given $\log(L_B)$ and plotted these residuals versus the fine-structure
\fsi.  Figure~\ref{fig:Dx} shows the result.  There is again an 
anticorrelation, similar in strength ($r=-0.44$) to the anticorrelation
of X-ray excess with \fsi.  Early-type galaxies with high fine-structure
index all lie {\it below} the mean $L_X$--$L_B$ relation, indicating that
they are hot-gas deficient.

Galaxies with detected \hi\ are shown as circles in Fig.~\ref{fig:LxLb} 
and Fig.~\ref{fig:Dx}. From these figures we see no trend of \hi\ detection 
probability with X-ray excess or with \fsi. \hi\ detections appear in all
types of galaxies plotted. The two merger remnants have \hi\ detected in tidal
tails (see Table~\ref{tab:SS92}).
The galaxy with the highest \fsi\ is the $\sim$0.5--0.7 Gyr old merger
remnant NGC 7252, which has $5\times 10^9 M_{\odot}$ of \hi\ in its tidal
tails, yet a dearth of \hi\ at its center (Hibbard \et\ 1994).
Table~\ref{tab:SS92} shows that there are a few other E and S0
galaxies with high fine-structure content
($3 < {\mit\Sigma} < 8$) that feature \hi\ in tidal tails or
outside the galaxy. From Table~\ref{tab:SS92} all instances of \hi\ in
rings or disks occur in galaxies of lower \fsi.  These galaxies cover a
range of X-ray excesses. Therefore we see a weak trend of \hi\ morphology
(but not of detection probability) with fine structure.  

\begin{figure*}
      \epsfig{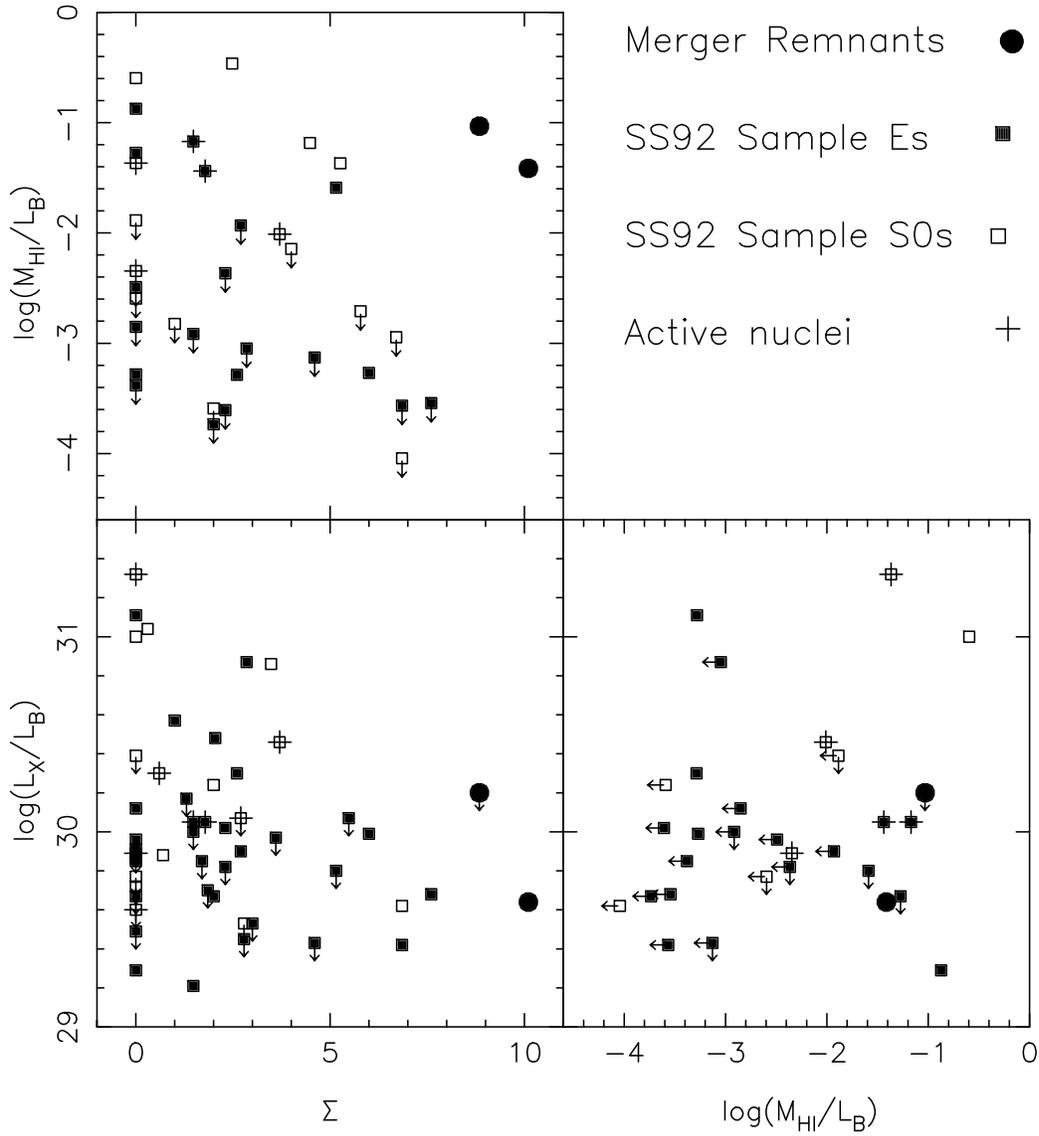}    
\caption{Composite plot showing cold- and hot-gas content versus \fsi.
Elliptical galaxies are shown as filled squares, S0 galaxies as open
squares, and the two recent merger remnants NGC 7252 and NGC 3921 as filled
circles.  Attached arrows indicate $3\sigma$ upper limits.  The three
panels show the parameters of cold-gas excess ($\log(M_{\rm HI}/L_B)$) 
(M$_{\odot}$ $L_{B_{\odot}}^{-1}$), hot-gas excess ($\log(L_X/L_B)$) 
(erg s$^{-1}$ $L_{B_{\odot}}^{-1}$), and fine-structure index (\fsi ) plotted
against each other. The X-ray data plotted include all SS92 galaxies
observed with ROSAT, as listed in Table~\ref{tab:SS92}.  Note the apparent
anticorrelation between $\log(L_X/L_B)$ and \fsi.}
\label{fig:Composite}
\end{figure*}

In Fig.~\ref{fig:Composite} we plot both the X-ray excess and the
measured \hi\ excess versus the fine-structure index.  Data from all the
galaxies listed in Table~\ref{tab:SS92} are included in this plot, with
data points coded by galaxy type.  Data points representing
ellipticals and S0 galaxies 
from the sample of SS92 are drawn as filled and open squares, respectively. 
The merger remnants NGC 7252 and NGC 3921 are represented by filled circles.
Galaxies with AGNs (from V\'eron-Cetty \& V\'eron 1996) are plotted as plus
signs.  Apart from the Seyfert galaxy NGC 3998, which has the highest X-ray
excess of our sample, the other four galaxies with AGNs (NGC 2768, 3032,
4278, 5273) and three LINERS (NGC 1052, 3226, 4036) are not extreme cases.
Therefore, the effect of X-ray emission from active nuclei appears to be
relatively small in these plots.

The top plot shows no clear trend of atomic gas with increasing \fsi\,.
In fact the ellipticals and S0s with the highest \fsi\ lie much lower in
this plot than the two merger remnants, indicating a dearth of \hi\,.
This is contrary to the naive expectation that recent merger remnants might
have copious amounts of tidally ejected \hi\ present in their
outer regions, gas that then gradually falls back into the remnant's body
(Hibbard \& Mihos 1995). We address this issue in more detail in
Hibbard \& Sansom (2000) where we discuss the possibilities that some 
high-\fsi\ systems may not be recent ($<$ few Gyr old) merger remnants of 
{\it gas-rich} galaxies;
\hi\ tidal tails may not survive for more than a few gigayears, or the
survivability of tidal \hi\ may depend on other factors.

In the two lower panels of Fig.~\ref{fig:Composite} we plot $\log(L_X/L_B)$
against both \fsi\ and $\log(M_{\rm HI}/L_B)$.  The plot of X-ray excess
against neutral-gas excess shows no discernible trend---there
are X-ray luminous galaxies both with large amounts of \hi\ and with
stringent limits on \hi.  This shows that there exists no clear
relationship between the cold and hot gas phases for this sample of
galaxies.  The lack of \hi\ and the low X-ray excess measured in relaxed,
early-type galaxies with high \fsi\ (see Table~\ref{tab:SS92})
imply a distinct lack of cold {\it and} hot gas in these suspected late
merger remnants.  Yet, gas is expected to have fallen back in from the tidal
tails. If these galaxies are 
the ancient remnants of disk-disk mergers then it is not clear where 
the cold gas formerly associated with the progenitors is now.  It may 
have been ionized by the star formation associated with the merger,
efficiently formed into stars, cooled to molecular form, or lost from the 
system. In major mergers a fair amount of the gas is known to turn into
stars, partially via the central-disk phenomenon (Schweizer 1998, esp.\
\S~5.3).  In future work we will investigate the molecular-gas content of
this sample to check if they have unusually large quantities of molecular 
gas when compared to early-type galaxies as a class (Roberts \et\ 1991).

We also looked for environmental effects on the X-ray excess of the sample
galaxies.  We find no correlations between the X-ray excess and Tully's
(1988) local-galaxy-density parameter $\rho$.  Most of the galaxies in
this sample are field objects or members of small groups.  Five galaxies
are Virgo Cluster members (marked by a `V' in Table~\ref{tab:SS92}).
These five galaxies cover a range of \fsi\ and X-ray excess. They are
not all X-ray luminous.  Therefore we find no evidence for any
environmental effect on the X-ray excess or its anticorrelation with
\fsi\ in this sample.

Figure~\ref{fig:Composite} shows that there are no clear trends between the 
hot- and cold-gas contents of these galaxies.  However, there is a clear
anticorrelation between the X-ray excess and the
amount of fine structure. As the lower left panel of Fig.~\ref{fig:Composite}
shows, there is a remarkable lack of galaxies with high \fsi\ and large
X-ray excess. We find that the ROSAT survey data of Beuing \et\ also
display this trend. This anticorrelation was discovered for two high fine
structure galaxies by Fabbiano \& Schweizer (1995), further quantified by
Mackie \& Fabbiano (1997), and is discussed by Schweizer (1998, esp.\
p.\ 198).  Peculiar E and S0 galaxies with high fine-structure content tend
to be X-ray weak, consistent with X-ray emission from stellar sources alone.
For example, $\log(L_X/L_B)$ is 29.54 (in the 0.5--2 keV band) for the bulge
of M31, with nearly equal contributions from the hard and soft emission
from low-mass X-ray binaries (Irwin \& Sarazin 1998). The anticorrelation
between fine structure and X-ray excess found in this work
supports the suggestion made by Fabbiano \& Schweizer (1995) that merger
remnants are X-ray weak. These authors suggest that a merger-induced
starburst drives a galactic superwind, which clears the remnant of
most hot gas.  After the starburst subsides, the hot gas is gradually
replenished by stellar mass-loss (Sandage 1957; Faber \& Gallagher 1976;
Forman \et\ 1985; Sarazin 1997) and perhaps to some small extent also by
the thermalization of returning tidal debris (Hibbard \& van Gorkom 1996),
although the $\log(L_X/L_B)$ vs $\log(M_{\rm HI}/L_B)$ plot argues against
the latter being the primary source of X-ray halos.

This suggests either that the time scale for creating hot halos in
early-type galaxies is long (a few Gyr), regardless of the cold-gas content
of the progenitors, or that E/S0 galaxies made in different ways have
different X-ray properties, with merger remnants forming a distinguishable,
X-ray underluminous class.  The range of $\log(L_X/L_B)$ values
in Fig.~\ref{fig:LxLb} suggests that the long time scale for halo production
is more likely: we do not see a dichotomy of behaviour but, rather, a gradual
change of X-ray excess with \fsi\,.

\section{Summary}
\label{sec:summary}

We have searched for correlations between the cold-gas content,
hot-gas content, and fine-structure index \fsi\ of early-type galaxies.

We find no correlation of \hi\ content or \hi\ detection probability with 
\fsi.  In particular we do not observe any tendency for 
high-\fsi\ systems to have large amounts of tidal \hi.
Also, the \hi\ content does not correlate with the
hot-gas content as measured by the X-ray excess.

However, we do observe an anticorrelation between \fsi\ and X-ray excess.
We find that E and S0 galaxies with high fine-structure content---thought
to be remnants of ancient mergers (``King Gap galaxies'')---have low X-ray 
luminosities.
This indicates that these galaxies, when compared to early-type galaxies
with little or no fine structure, are deficient in hot gas.
During mergers, X-ray emission from hot gas is seen to rise (with evidence 
for outflows) and then fall after a relatively short time (Read \& Ponman 
1998). This and the current results are consistent with hot halos forming 
in early-type galaxies over a time scale of several Gyrs.  Hence, such
halos tend to appear {\it after} the morphological disturbances
from past mergers have faded away.  If one excludes galaxies with AGNs, only
normal E and S0 galaxies with {\it low} levels of fine structure (i.e.,
relaxed systems) are strong X-ray emitters, with emission well in excess of
the contribution expected from stars.

\acknowledgments

We thank Jacqueline van Gorkom and David Schiminovich for permission to
refer to their results ahead of publication, and Keith Arnaud for the use of
his code for hydrogen column densities.
This research has made use of data obtained from the Leicester Database 
and Archive Service at the department of Physics and Astronomy, Leicester 
University, UK. 
One of us (F.S.) gratefully acknowledges partial support from NSF through
Grants AST\,95-29263 and AST\,99-00742.

\small

\begin{deluxetable}{lrccrccc}
\tablecaption{ROSAT PSPC data from Pointed Observations.}
\tablehead{
\colhead{System} &
\colhead{\hfil Source } &
\colhead{\hfil Error } &
\colhead{R} &
\colhead{Exposure} &
\colhead{$N_{\rm HI}$} &
\colhead{Flux} &
\colhead{Error} \\
& \colhead{Counts} & 
\colhead{Counts} & 
\colhead{'} &
\colhead{sec} &
\colhead{$\times 10^{20}$ cm$^{-2}$} &
\colhead{erg cm$^{-2}$ s$^{-1}$} &
\colhead{erg cm$^{-2}$ s$^{-1}$} \\
}\startdata
$R < 20\arcmin$ & & & & & & & \\
NGC 524  &  268.3 & 18.8 &2 &11171 &4.53 & 0.300E-12  & 0.601E-13 \\
NGC 596  & $<$23.6 & 10.3 &2 & 4055 &2.46 & $<$0.646E-13 & 0.292E-13 \\
NGC 1052 &  474.9 & 27.7 &2 &13975 &3.11 & 0.400E-12 & 0.743E-13 \\
NGC 2300 & 1729.1 & 55.2 &5 &17446 &6.39 & 0.142E-11 & 0.289E-12 \\
NGC 2768 &  177.9 & 20.3 &4 & 4766 &4.24 & 0.466E-12 & 0.102E-12 \\
NGC 3193 &   67.7 & 13.3 &2 & 4617 &3.39 & 0.172E-12 & 0.455E-13 \\
NGC 3226 &  745.0 & 38.2 &3 &19547 &3.54 & 0.448E-12 & 0.824E-13 \\
NGC 3605 &  116.7 & 15.8 &1 &23853 &2.88 & 0.576E-13 & 0.128E-13 \\
NGC 3607 & 1637.9 & 68.3 &4 &23853 &2.89 & 0.808E-12  & 0.146E-12 \\
NGC 3608 &  500.1 & 32.0 &2 &23853 &2.89 & 0.247E-12  & 0.463E-13 \\
NGC 3610 &  146.6 & 20.6 &2 &10784 &0.83 & 0.129E-12 & 0.203E-13 \\
NGC 3640 &  125.9 & 19.4 &2 &15107 &3.23 & 0.980E-13 & 0.230E-13 \\
NGC 3998 &38927.1 &205.9 &3 &60722 &1.32 & 0.622E-11  & 0.605E-12 \\
NGC 4125 &  406.5 & 31.1 &3 & 5707 &1.45 & 0.692E-12 & 0.855E-13 \\
NGC 4203 & 5345.7 & 74.4 &1.5 &22663 &1.68 & 0.248E-11  & 0.264E-12 \\
NGC 4261 & 2710.9 & 85.1 &4 &21893 &1.99 & 0.133E-11  & 0.149E-12 \\
NGC 4278 &  173.3 & 16.5 &2 & 3411 &1.88 & 0.535E-12 & 0.760E-13 \\
NGC 4283 & $<$19.9 & 8.7 &2 & 3411 &1.88 & $<$0.627E-13  & 0.281E-13 \\
NGC 4374 & 4054.3 & 73.4 &2 &33659 &1.75 & 0.126E-11  & 0.953E-13 \\
NGC 4382 &  411.2 & 50.2 &4 & 8495 &2.37 & 0.538E-12 & 0.888E-13 \\
NGC 4552 & 1842.7 & 49.8 &2 &16660 &1.68 & 0.116E-11 & 0.127E-12 \\
NGC 4697 & 3359.1 &128.0 &5 &45235 &2.73 & 0.825E-12  & 0.969E-13 \\
NGC 5273 &  152.8 & 16.8 &2 & 5386 &1.12 & 0.275E-12  & 0.403E-13 \\
NGC 5322 &  700.5 & 46.2 &2 &34778 &1.73 & 0.212E-12 & 0.263E-13 \\
NGC 5846 & 3803.7 & 76.0 &5 & 8804 &3.96 & 0.540E-11  & 0.102E-11 \\
NGC 7252 &  133.2 & 23.2 &2 &17371 &2.28 & 0.852E-13 & 0.176E-13 \\
NGC 7619 & 3548.5 & 97.3 &7 &18221 &5.35 & 0.243E-11 & 0.461E-12 \\
NGC 7626 &  515.1 & 29.4 &2 &18221 &5.35 & 0.353E-12 & 0.693E-13 \\
$20\arcmin < R < 60\arcmin$ & & & & & & & \\
NGC 584  & $<$43.0 & 18.7 &3 & 4055 &2.49 & $<$0.118E-12 & 0.528E-13 \\
NGC 3032 & $<$90.4 & 47.0 &3 &12206 &2.73 & $<$0.823E-13 & 0.437E-13 \\
NGC 3065 & 339.6   & 41.2 &3 &10145 &2.52 & 0.372E-12 & 0.612E-13 \\
NGC 3921 & $<$49.1 & 21.3 &3 & 3851 &1.17 & $<$0.124E-12 & 0.551E-13 \\
NGC 4168 & $<$84.7 & 36.5 &3 &14300 &2.01 & $<$0.637E-13 & 0.283E-13 \\
NGC 4660 & $<$54.5 & 23.2 &3 & 4138 &2.12 & $<$0.146E-12 & 0.644E-13 \\
NGC 4915 & $<$63.7 & 23.0 &3 & 7429 &2.87 & $<$0.101E-12 & 0.405E-13 \\
NGC 5198 & $<$123.2 & 53.0 &3 &23956 &1.73 & $<$0.541E-13 & 0.240E-13 \\
NGC 5557 & $<$92.2 & 39.7 &3 &14242 &1.09 & $<$0.629E-13 & 0.277E-13 \\
NGC 5831 & $<$60.0 & 26.1 &3 & 5941 &4.37 & $<$0.126E-12 & 0.598E-13 \\
\enddata
\label{tab:ROSAT}
\end{deluxetable}

\scriptsize

\begin{deluxetable}{lccccclc}
\tablecaption{Galaxies from SS92 with \hi\ Mapping and/or ROSAT X-ray 
Observations}
\tablehead{
\colhead{System} &
\colhead{${\mit\Sigma}$} &
\colhead{$B$} &
\colhead{D} &
\colhead{$\log(M_{\rm HI})$} &
\colhead{\hi\ Morphology} &
\colhead{\hi\ Reference} &
\colhead{$\log(L_X/L_B)$} \\
& & mag & Mpc & $M_\odot$ & & & erg s$^{-1}$L$_{B_{\odot}}^{-1}$ 
}\startdata
NGC 7252 & 10.1 & 12.06 & 63.5 & 9.72 & Tidal Tails & Hibbard \et\ 1994 & 29.64 \\ 
NGC 3921 & 8.84 & 13.06  & 77.9 & 9.56 & Tidal Tails & Hibbard \& van Gorkom 1996 & $<$30.20  \\
NGC 3610 & 7.60 & 11.70 & 29.2 & $<$6.90 & n/d & Hibbard \& Sansom 2000 & 29.68 \\ 
NGC 3640 & 6.85 & 11.36 & 24.2 & $<$6.85 & n/d & Hibbard \& Sansom 2000 & 29.42 \\ 
NGC 4382 V & 6.85 & 10.00 & 16.8 & $<$6.60 & n/d & Hibbard \& Sansom 2000 & 29.62 \\ 
NGC 7585 & 6.70 & 12.33 & 46.0 & $<$7.64 & n/d & Schiminovich
\et\ 2000 & --- \\
NGC 4125 & 6.00 & 10.65 & 24.2 & 7.43 & Outside Body & Rupen, personal comm. &29.99 \\  
NGC 7600 & 5.78 & 12.91 & 45.8 & $<$7.64 & n/d & Schiminovich \et\ 2000 & --- \\
NGC 4915 & 5.48 & 12.95 & 42.0 & --- & --- & --- & $<$30.07 \\
NGC 0474 & 5.26 & 12.37 & 32.5 & 8.90 & Tidal Debris & Schiminovich \et\ 1997 & --- \\
NGC 5018 & 5.15 & 11.69 & 40.9 & 9.15 & Bridge & Kim \et\ 1988  & $<$29.80B \\
NGC 0596 & 4.60 & 11.84 & 23.8 & $<$7.08 & n/d & Schiminovich \et\ 2000 
 & $<$29.43 \\
NGC 2911 & 4.48 & 12.50 & 41.8 & 9.25 & unknown & Chamaraux \et\ 1987 & --- \\
NGC 7332 & 4.00 & 12.02 & 18.2 & $<$7.76 & n/d & Haynes 1981 & --- \\
NGC 3226 & 3.70 & 12.30 & 23.4 & $\sim$8.0 & near Tidal Tails & Mundel \et\ 1995 & 30.46 \\
NGC 5831 & 3.60 & 12.45 & 28.5 & --- & --- & --- & $<$29.97 \\
NGC 3065 & 3.48 & 13.50 & 31.3 & --- & --- & --- & 30.86 \\
NGC 4168 V & 3.00 & 12.11 & 16.8 & --- & --- & --- & $<$29.53 \\
NGC 2300 & 2.85 & 12.07 & 31.0 & $<$7.30 & n/d & Schiminovich \et\ 2000 & 30.87 \\ 
NGC 0584 & 2.78 & 11.44 & 23.4 & --- & --- & --- & $<$29.53 \\
NGC 5557 & 2.78 & 11.92 & 42.6 & --- & --- & --- & $<$29.45 \\
NGC 3032 & 2.70 & 13.18 & 24.5 & --- & --- & --- & $<$30.07 \\
NGC 3605 & 2.70 & 13.13 & 16.8 & $<$7.46 & n/d & Haynes 1981 & 29.90 \\ 
NGC 7626 & 2.60 & 12.16 & 45.6 & 7.36 & Outside Body & Hibbard \& Sansom 2000 & 30.30 \\  
NGC 2685 & 2.48 & 12.12 & 16.2 & 9.30 & Polar Ring & Shane 1980 & --- \\
NGC 4374 V & 2.30 & 10.09 & 16.8 & $<$7.0 & n/d & Schiminovich
\et\ 2000 & 30.02 \\
NGC 5576 & 2.30 & 11.85 & 26.4 & $<$7.93 & n/d & Haynes 1981 & $<$29.82B \\
NGC 5982 & 2.04 & 12.04 & 38.7 & --- & --- & --- & 30.48B \\
NGC 4552 V & 2.00 & 10.73 & 16.8 & $<$6.76 & n/d & Schiminovich \et\ 2000 & 30.24 \\ 
NGC 5322 & 2.00 & 11.14 & 31.6 & $<$7.00 & n/d & Hibbard \& Sansom 2000 & 29.67 \\ 
NGC 5198 & 1.85 & 12.69 & 39.0 & --- & --- & --- & $<$29.70 \\
NGC 1052 & 1.78 & 11.41 & 17.8 & 8.69 & Tidal Tails & van Gorkom \et\ 1986 & 30.05 \\  
NGC 3156 & 1.70 & 13.07 & 18.6 & --- & --- & --- & $<$29.85B \\
NGC 0636 & 1.48 & 12.41 & 24.2 & $<$7.08 & n/d & Schiminovich \et\ 2000 &$<$30.00B \\
NGC 3377 & 1.48 & 11.24 & 8.1 & --- & --- & --- & \,\,\,\,\, 29.21BB \\
NGC 4278 & 1.48 & 11.09 &  9.7 & 8.56 & Disk & Raimond \et\ 1981 & 30.05 \\  
NGC 3818 & 1.30 & 12.67 & 25.1 & --- & --- & --- & $<$30.17B \\
NGC 3245 & 1.00 & 11.70 & 22.2 & $<$7.38 & n/d & Chamaraux \et\ 1987 & --- \\
NGC 4261 & 1.00 & 11.41 & 35.1 & --- & --- & --- & 30.57 \\
NGC 0524 & 0.70 & 11.30 & 32.1 & --- & --- & --- & 29.88 \\
NGC 5273 & 0.60 & 12.44 & 21.3 & --- & --- & --- & 30.30 \\
NGC 5846 & 0.30 & 11.05 & 28.5 & --- & --- & --- & 31.04 \\
NGC 2768 & 0.00 & 10.84 & 23.7 & 8.26 & Outside Body & Dijkstra \et\ 2000 
& 29.89 \\ 
NGC 2974 & 0.00 & 11.87 & 28.5 & 9.08 & Disk & Kim \et\ 1988 & $<$29.67B \\
NGC 3193 & 0.00 & 11.83 & 23.2 & $<$6.81 & n/d & Williams \et\ 1991 & 29.85 \\
NGC 3379 & 0.00 & 10.24 &  8.1 & 9.04 & 200 kpc Ring & Schneider 1989 & 
\,\,\,\,\, 29.29BB \\
NGC 3607 & 0.00 & 10.82 & 19.9 & $<$7.61 & n/d & Haynes 1981 & 30.12 \\
NGC 3608 & 0.00 & 11.70 & 23.4 & $<$7.76 & n/d & Haynes 1981 & 29.96 \\
NGC 3998 & 0.00 & 11.61 & 21.6 & 8.85 & Polar Ring & Knapp \et\ 1985b & 31.32 
\\
NGC 4036 & 0.00 & 11.57 & 24.6 & --- & --- & --- & $<$29.60B \\
NGC 4203 & 0.00 & 11.80 &  9.7 & 8.85 & Ring & van Driel \& van Woerden 1991 & 31.00 \\
NGC 4283 & 0.00 & 13.00 &  9.7 & --- & --- & --- & $<$29.88 \\
NGC 4589 & 0.00 & 11.69 & 30.0 & --- & --- & --- & $<$29.49B \\
NGC 4660 V & 0.00 & 12.16 & 16.8 & --- & --- & --- & $<$29.91 \\
NGC 4697 & 0.00 & 10.14 & 23.3 & --- & --- & --- & 29.86 \\
NGC 5485 & 0.00 & 12.31 & 32.8 & --- & --- & --- & $<$29.72B \\
NGC 5574 & 0.00 & 13.23 & 28.7 & $<$7.93 & n/d & Haynes 1981 & $<$30.39B \\
NGC 7457 & 0.00 & 12.09 & 12.3 & $<$6.94 & n/d & Chamaraux \et\ 1987 
& $<$29.77B \\
NGC 7619 & 0.00 & 12.10 & 50.7 & 7.48 & Outside Body & Dijkstra \et\ 2000 
& 31.11 \\  
\enddata
\tablenotetext{B}{Ratio from Beuing \et\ (1999), using ROSAT survey data.}
\tablenotetext{BB}{Ratio from Brown \& Bregman (1998), using ROSAT HRI 
data.}
\label{tab:SS92}
\end{deluxetable}

\end{document}